\documentstyle[prl,aps,preprint]{revtex}
\begin{document}

\draft

\title{The spin structure of the proton in a
non--relativistic quark model}

\author{J. Keppler}

\address{Institut f\"ur theoretische Physik III, Universit\"at
Erlangen--N\"urnberg, \\ Staudtstra\ss e 7, 91058 Erlangen, Germany}

\author{H. M. Hofmann}

\address{Institut f\"ur theoretische Physik III, Universit\"at
Erlangen--N\"urnberg, \\ Staudtstra\ss e 7, 91058 Erlangen, Germany \\
$\text{and}$ \\
European Center for Theoretical Studies in Nuclear Physics
and Related Areas $(ECT^{*})$, \\
Villa Tambosi, Via delle Tabarelle 286,
I--38050 Villazzano (Trento), Italy}

\date{\today}

\maketitle
\begin{abstract}
The spin structure of the proton is investigated in the framework of
an extended quark potential model which in addition to the conventional
$3q$--structure also takes into account $(3q)(q\bar q)$--admixtures in
the proton wave function. For reasons of parity such admixtures contain
an odd orbital angular momentum thus allowing the proton spin to be shared
among quark spins and orbital angular momenta. We show that only certain
admixtures are suited for a significant reduction of the quark spin
content of the proton as suggested by the EMC--result. Within a Hamiltonian
model quark spin contributions to the proton spin
down to $0.5$ can be reproduced easily .
\end{abstract}

\pacs{}

\narrowtext

In 1988 the European Muon Collaboration (EMC) \cite{ref1,ref2} determined
the spindependent proton structure function $g^{p}_{1}(x)$ down to values
of $x$ as small as $0.015$. Furthermore an estimate for the extrapolation to
$x \to 0$ was presented, thus being able to give a first result for the
integral of $g^{p}_{1}(x)$. An evaluation of the underlying
deep inelastic scattering process in the framework of the naive quark
parton model results in associating the quantity
$\int^{1}_{0}dx \, g^{p}_{1}(x)$ with a linear combination of the quark
spin contributions to the proton spin. Applying additional information
from neutron and hyperon $\beta$--decay it is then possible to analyze the
spin structure of the proton. Such a straightforward
analysis leads, however, to the surprising conclusion that the
contribution of all the quarks to the nucleon spin is unexpectedly small
and compatible with zero \cite{ref3,ref4}. Accordingly a huge fraction of the
proton spin has to be carried by gluonic degrees of freedom or internal
orbital angular momenta.

The consequence for the nonrelativistic quark model and a possible way out of
the ''spin crisis'' was first discussed by Jaffe and Lipkin \cite{ref5}
who modified the proton wave function in admixing $(3q)(q\bar q)$--components
to the conventional $3q$--object. This can be expressed as
\begin{equation}
|\Psi^{\text{tot}}_{p}\rangle = \nu_{0}\, |\Psi_{p}(3q)\rangle +
\sum_{i=1} \nu_{i}\, |\Psi_{i}[(3q)(q\bar q)]\rangle \:.
\label{eq1}
\end{equation}
The structure of such $(3q)(q\bar q)$--components which can be
considered as baryon--meson--combinations is depicted in Fig.\
\ref{fig1}.
Jaffe and Lipkin \cite{ref5}
took into account just two $5q$--components in which
the $3q$--substructure was assumed to have the quantum numbers of the
proton and the odd orbital angular momentum which has to be introduced
for reasons of parity was used to construct P--wave vector mesons, one of
them being coupled to $J_{M}=0$, the other to $J_{M}=1$. They found
that the free coefficients $\nu_{1,2}$ of the total wave function can be
chosen in a way to reproduce the weak coupling constants as well as a
quark spin content which is compatible with zero. In such a scenario
the proton spin is effectively carried by the orbital angular momentum
$L_{M}$, see Fig.\ \ref{fig1},
which in a non--relativistic quark model
is apart from the quark spins the only source of the nucleon spin.

A closer look at the
Jaffe--Lipkin--ansatz \cite{ref5}, however, reveals several
problems. First the choice of the proton
wave function seems to be quite arbitrary and cannot be justified by
physical arguments. The $q\bar q$--pairs taken into account are
P--wave vector mesons while on energetic grounds one would expect the
$5q$--component to be dominated by pseudoscalar
S--wave mesons, like the pion.
Second the fit of the coefficients leads to a complete suppression
of the $3q$--structure $|\Psi_{p}(3q)\rangle$ and a dominance of the
$5q$--admixtures, a scenario which seems rather questionable.
Finally the model does not specify the interaction
generating the $q\bar q$--pairs
which may be regarded as effective sea quarks.
This also prevents studying
the behaviour of other important properties of the nucleon like mass
and magnetic moment.

Nevertheless the simple model proposed by
Jaffe and Lipkin \cite{ref5} serves as
a guideline to perform an investigation of the spin structure of
the proton in an extended quark potential model \cite{ref6a,ref6}.
In this model the
interactions of the quarks are mediated by one gluon exchange potentials
which because of the sea quark admixtures also include an annihilation
as well as pair creation and pair destruction potentials. Together with
the operators for kinetic energy and confinement the sum of these
potentials constitutes the Hamiltonian for the proton state in the
extended non--relativistic quark model. For details of the interaction
see reference \cite{ref6}. The free parameters of this
Hamiltonian are the quark masses $m_{u}=m_{d}$ and $m_{s}$,
the strong coupling constant $\alpha_{s}$,
the confinement strength $a$, and a
cutoff parameter $\Lambda$ which enters a form factor \cite{ref7a} of the
quark--gluon--vertex. The latter is introduced in order to regularize
the potential and make variational calculations possible. The approximate
solution of the Schr\"odinger equation is then given by the
Refined Resonating Group Method \cite{ref7}. As an ansatz for the wave
function we start from Eq.\ (\ref{eq1}) allowing $5q$--components which
contain either S-- or P--wave mesons
and properly antisymmetrize
the quarks of the total wave function. In the case
of S--wave mesons the odd
orbital angular momentum is on the relative motion
between the baryon and the meson,i.e.
$L_{\text{rel}}=1$, in the other case $L_{\text{rel}}=0$
and $L_{M}=1$.
Admixtures with more than five particles, orbital angular momenta of three,
or more complex angular momentum configurations
are energetically suppressed and therefore neglected.
Altogether 66
baryon--meson--combinations which can be coupled to the quantum numbers
of the proton are taken into account, among them also ones containing strange
quarks and color octet fragments. They are called channels and
determined by the following coupling scheme in spin, isospin or color
space, see Fig.\ \ref{fig1} for the obvious notation,
\begin{eqnarray}
[J_{p}T_{p}C_{p}] & = & \biggl[ \Bigl[ J_{B}\times [S_{M}\times
L_{M}]J_{M} \Bigl]S_{c} \times L_{\text{rel}} \biggl] J_{p} \nonumber  \\
& & \mbox{} \otimes \,[T_{B}\times T_{M}]T_{p} \,\otimes\,
[C_{B}\times C_{M}]C_{p} \:,
\label{eq2}
\end{eqnarray}
where $S_{c}$ is the channel spin. The coefficients $\nu_{i}$
of the individual channels are
obtained by diagonalizing the Hamiltonian matrix and are therefore {\em no\/}
free parameters in our model.

Before starting the numerical calculations we study the spin properties of
individual channels. These investigations reveal that only certain classes
of $5q$--components meet the requirements for an appreciable reduction
of the quark spin contribution.
To keep the notation simple, we classify only those channels which will
be discussed later on. For channels containing S--wave mesons it is
sufficient to specify the channel spin $S_{c}$, therefore we denote the
channels by $|S_{c};\text{S}\rangle$. Of the channels containing
P--wave mesons only those play a special role where the baryon has
the quantum number
$J_{B}=1/2$, but the spin $S_{M}$ and the total angular momentum $J_{M}$
have to be specified, hence, we denote them by
$|S_{M},J_{M};\text{P}\rangle$. On the one hand the quark spin contribution
to the nucleon spin can be reduced by channels which
have a negative total spin content of the
quarks. Into this class, in the following designated as class--I,
fall all channels with $|1/2;\text{S}\rangle$
and $|0,1;\text{P}\rangle$.
Examples for the former and latter category are N$\pi$, N$\rho$,
N$\omega$, N$\eta$, $\Delta\rho$, and
N$\text{b}_{1}(1235)$, N$\text{h}_{1}(1170)$ \cite{ref8a}
respectively. It is straightforward to show that
\begin{eqnarray}
& & \langle 1/2;\text{S}|\sigma_{z}|
1/2;\text{S}\rangle \nonumber \\
& = & \langle 0,1;\text{P}|\sigma_{z}|
0,1;\text{P}\rangle
=  -\frac{1}{3} \:,
\label{eq3}
\end{eqnarray}
holds. Since different channels of this class
do not mix via the spin operator, the admixture of
such $5q$--structures to the conventional $3q$--structure will therefore
in any case result in a reduction of the quark spin content
$\Sigma(\Psi_{p}^{\text{tot}}) \equiv
\langle \Psi_{p}^{\text{tot}} |\sigma_{z}| \Psi_{p}^{\text{tot}}\rangle$
of the physical proton. All remaining channels
are characterized by a positive quark spin contribution.
The second possibility to reduce $\Sigma(\Psi_{p}^{\text{tot}})$ then
consists in admixing channels which are coupled strongly via the
spin operator. Into this class, in the following designated as class--II,
fall channels containing
P--wave vector mesons
differing by the $J_{M}$--quantum number only. Examples for such pairs are
$\bigl(\text{Na}_{0}(980),\text{Na}_{1}(1260)\bigr)$
and $\bigl(\text{Nf}_{0}(975),\text{Nf}_{1}(1285)\bigr)$
\cite{ref8a}.
A simple calculation yields
\begin{eqnarray}
\langle 1,0;\text{P}|\sigma_{z}|
1,0;\text{P}\rangle & = & 1 \:,   \\         \label{eq4}
\langle 1,1;\text{P}|\sigma_{z}|
1,1;\text{P}\rangle & = & \frac{1}{3} \:,    \\   \label{eq5}
\langle 1,0;\text{P}|\sigma_{z}|
1,1;\text{P}\rangle & = & -\frac{2\sqrt{2}}{3} \:.
\label{eq6}
\end{eqnarray}
Incorporating one of these channel doublets into the ansatz
of their wave function, Jaffe and Lipkin \cite{ref5}
took a very suitable choice and
succeeded in fitting the coefficients to reproduce the $\beta$--dacay as
well as the EMC--data. Performing model calculations based on a Hamiltonian
to determine the nucleon wave function, it is quite evident, however, that
the corresponding expansion coefficients of these channel pairs are no fit
parameters and therefore play a crucial role. If they are too small or
have the wrong sign no reduction of the total quark spin content can be
expected.

Having analyzed the spin properties of individual channels, we proceed to
investigate the influence of several relevant channels on the spin structure
of the proton. As a first step we determine the maximal
strength of the $3q$--structure in the nucleon wave function which
does not contradict the experimental
finding for the quark spin content. Admixing an arbitrary
number of channels out of the
$|1/2;\text{S}\rangle$-- or  $|0,1;\text{P}\rangle$--sectors, one
simply gets
\begin{equation}
\Sigma(\Psi_{p}^{\text{tot}}) = \nu_{0}^{2}-\frac{1}{3}\sum_{i=1}
\nu_{i}^{2} = \frac{4}{3}\nu_{0}^{2}- \frac{1}{3} \:,
\label{eq7}
\end{equation}
using $\sum_{i=0} \nu_{i}^{2} = 1$. Thus,
in this case the quark spin content depends on the probability
amplitude of the $3q$--component only. According to the previous paragraph
a second promising ansatz consists in including channel doublets with
a large negative spin overlap,
those with a positive one do not exist.
This results in a more complex expression
for  $\Sigma(\Psi_{p}^{\text{tot}})$ which in addition to $\nu_{0}$ also
involves the coefficients of the $5q$--components. From such expressions
we altogether find that even after restricting to the most favourable
channel combinations and an optimal choice of the various coefficients
a value of $\Sigma(\Psi_{p}^{\text{tot}}) \approx 0$  can be achieved only
if $\nu_{0}^{2} \leq 0.3$. Taking into account the possibility of
$\Sigma(\Psi_{p}^{\text{tot}}) \approx 0.5$, the squared amplitude of the
$3q$--structure has to be less than $\approx 0.7$.

After having studied the fundamental connection between the
basic channels and the quark spin content of the proton at an analytic
level it is now of central interest which coefficients and which
spin structure
are realized in the extended quark potential model. As an appropriate
starting point we employ a previously given parameter set \cite{ref8},
with $m_{u}=m_{d}= 330\, \text{MeV}$, $\alpha_{s}=1.7$,
$a=-19\, \text{MeV}/\text{fm}^{2}$, extended by $m_{s}= 600\, \text{MeV}$
and $\Lambda= 300\, \text{MeV}$. The calculations yield that solely the
admixture of channels out of the
$|1/2;\text{S}\rangle$-- and  $|0,1;\text{P}\rangle$--sectors
leads to a reduction of the total quark spin content of the proton.
In this case the probability amplitude of the $3q$--structure is
$\nu_{0}^{2} \approx 0.85$ which corresponds to
$\Sigma(\Psi_{p}^{\text{tot}}) \approx 0.80$ according to eq.\ (\ref{eq7}).
It also appears that the essential contribution to this effect stems from
the five most dominant class--I channels
N$\pi$, N$\rho$, N$\omega$, N$\eta$, and $\Delta\rho$.
These are therefore the relevant
candidates for further considerations. Channel doublets with a large
negative spin overlap, such as
(N$\text{a}_{0}$,N$\text{a}_{1}$) and
(N$\text{f}_{0}$,N$\text{f}_{1}$) out of
class--2, couple comparatively well to the $3q$--component. The value
of $\Sigma(\Psi_{p}^{\text{tot}})$, however, even increases with respect
to the one of the $3q$--proton. This means that the corresponding expansion
coefficients resulting from our quark model calculations are {\em not\/}
suited for decreasing the quark spin content at all.

Finally we would like to study whether and to which degree a further
reduction of $\Sigma(\Psi_{p}^{\text{tot}})$ is obtainable within the limits
of an acceptable variation of the parameter set which also reproduces
fundamental properties of the nucleon like mass $m_{p}$, magnetic moment
$\mu_{p}$, and the $\beta$--decay constant $(g_{A}/g_{V})_{\text{np}}$ of
the neutron. Since we restrict ourselves to the dominant $5q$--structures
and therefore omit channels with strange quarks,
the parameters remaining for a variation are
$m_{q} \equiv m_{u}=m_{d}$, $\alpha_{s}$, and $a$. The cutoff is kept fixed
to the value of $\Lambda =  300\, \text{MeV}$. We find in the
case of class--2 channel admixtures parameter sets which
both reproduce the basic properties of the nucleon and lead to a significant
reduction of the $3q$--amplitude in the wave function. This, however,
does not lead to
a decrease of the quark spin content because parameter variations
do not change the unsuitable phases of the expansion coefficients. On the
other hand class--1 channel admixtures guarantee a reduction of the quark
spin in any case. We obtain $\Sigma(\Psi_{p}^{\text{tot}}) \approx 0.5$ by
fixing the parameters at
$m_{q}= 250\, \text{MeV}$, $\alpha_{s}=4.0$, and
$a=-170\, \text{MeV}/\text{fm}^{2}$. These values lie within the range of
already existing parameter sets in the literature \cite{ref8,ref9,ref10}.
We therefore believe that a parameter set leading to a proton configuration
in which quark spins and orbital angular momenta
share the nucleon spin equally is quite reasonable.
It remains to be mentioned that our calculation yields
in addition to the proton mass, the magnetic moment of the nucleon,
and the N--$\Delta$--mass splitting also a $\beta$--decay constant
$(g_{A}/g_{V})_{\text{np}} = 1.23$ which is in perfect agreement with
experiment. This means that the extended quark potential model cures
a well known shortcoming of the conventional non--relativistic quark
model, namely the overestimated ratio of the weak
coupling constants. Any further sizable reduction of the
quark spin content in the nucleon, however, would require an enormous
modification of the original parameter set which by no means seems to be
acceptable to us. Moreover, in this case the
agreement between the calculated
and the experimentally determined static properties of the nucleon
deteriorates.

In summary we have performed a careful analysis of individual $5q$--channels
with respect to their angular momentum structure. For reasons of parity
these channels contain an odd orbital angular momentum thus allowing the
proton spin to be shared among quark spins and orbital angular momenta.
We demonstrated that only certain
baryon--meson--combinations are in principle
suited for a reduction of the quark spin contribution to the nucleon spin.
These are either channels with a negative quark spin content or channel
doublets with a huge negative spin overlap. In the framework of consistent
quark model calculations we found, however, that an admixture of such
channel doublets does not lead to a reduction of the quark spin in the
nucleon. The success of the Jaffe--Lipkin--fit \cite{ref5} is therefore not
reproducable in an extended quark potential model. Choosing on the other
hand a wave function which in addition to the $3q$--structure consists
of the channels N$\pi$, N$\rho$, N$\omega$, N$\eta$ and $\Delta\rho$, one
can reproduce quark spin contributions down to $50\%$.
This result is the lower bound which can be achieved employing a reasonable
parameter set and clearly demonstrates the limitations of non--relativistic
quark models if the EMC--result should be confirmed in the future.

We would like to acknowledge helpful discussions with M. Unkelbach and
H. Berger.

\begin{figure}
\caption{The structure of $(3q)(q\bar q)$--admixtures in angular
momentum space.}
\label{fig1}
\end{figure}

\end{document}